\documentclass[twocolumn,notitlepage,aps,prb,10pt,superscriptaddress,floatfix,letterpaper
]{revtex4-1}

\usepackage{hyperref}
\usepackage[usenames,dvipsnames]{color}
\usepackage{amsmath}
\usepackage[english]{babel}
\usepackage{amssymb}
\usepackage{graphicx}
\usepackage{epstopdf}
\usepackage[normalem]{ulem}
\usepackage{subfigure}
\usepackage{todonotes}
\usepackage{braket}

\definecolor{linkcolor}{HTML}{399B03}
\definecolor{urlcolor}{HTML}{399B03}
\hypersetup{pdfstartview=FitH, linkcolor=linkcolor,urlcolor=urlcolor,colorlinks=true}
\hypersetup{
    unicode=false,          
    pdftoolbar=true,        
    pdfmenubar=true,        
    pdffitwindow=false,     
    pdfstartview={FitH},    
    pdfauthor={},         
    colorlinks=true,        
    linkcolor=blue,         
    citecolor=red,          
}

\newcommand{\li}{\mathbf{i}}
\newcommand{\lj}{\mathbf{j}}
\newcommand{\lk}{\mathbf{k}}
\newcommand{\ti}{\mathbf{\tilde{i}}}
\newcommand{\tj}{\mathbf{\tilde{j}}}
\newcommand{\tk}{\mathbf{\tilde{k}}}
\newcommand{\tq}{\mathbf{\tilde{q}}}
\newcommand{\cK}{\mathbf{K}}
\newcommand{\cQ}{\mathbf{Q}}
\newcommand{\cI}{\mathbf{I}}
\newcommand{\cJ}{\mathbf{J}}
\newcommand{\bigO}[1]{\mathcal{O}(#1)}

\begin{document}

\title{Momentum-Space Cluster Dual Fermion Method}

\author{Sergei Iskakov}
\affiliation{%
 Department of Physics, University of Michigan, Ann Arbor, Michigan 48109, USA
}%
\author{Hanna Terletska}
\affiliation{%
 Department of Physics, University of Michigan, Ann Arbor, Michigan 48109, USA
}%
\affiliation{%
 Department of Physics, Middle Tennessee State University, Murfreesboro, TN 37132, USA
}%
\author{Emanuel Gull}%
\affiliation{%
 Department of Physics, University of Michigan, Ann Arbor, Michigan 48109, USA
}%
\date{\today}

\begin{abstract}
Recent years have seen the development of two types of non-local extensions to the single-site dynamical mean field theory. On one hand,
cluster approximations, such as the dynamical cluster approximation, recover short-range momentum-dependent correlations non-perturbatively.
On the other hand, diagrammatic extensions, such as the dual fermion theory, recover long ranged corrections perturbatively. The correct
treatment of both strong short-ranged and weak long-ranged correlations within the same framework is therefore expected to lead to a quick
convergence of results, and offers the potential of obtaining smooth self-energies in non-perturbative regimes of phase space.
In this paper, we present an exact cluster dual fermion method based on an expansion around the dynamical cluster approximation. Unlike previous
formulations, our method does not employ a coarse graining approximation to the interaction, which we show to be the leading source of error
at high temperature, and converges to the exact result independent of the size of the underlying cluster. We illustrate the power of the
method with results for the second-order cluster dual fermion approximation to the single-particle self-energies and double occupancies.
\end{abstract}

\maketitle

\section{Introduction}
The complexity of the exact numerical solution of fermionic lattice model systems such as the two-dimensional Hubbard model\cite{PhysRevX.5.041041} is
generally believed to scale exponentially as a function of system size. There is therefore a need for approximations that capture the
salient aspects of these models while remaining computationally affordable, such that routine simulations of large parameter spaces can be
performed.

In particular, there is  a need for reliable approximations to single- and two-particle response functions for `correlated' systems at
intermediate-to-large interaction strengths and at finite temperature, as these correspond to the quantities measured in experiments such as
angle-resolved photoemission spectroscopy,\cite{RevModPhys.75.473} scanning tunneling
microscopy,\cite{RevModPhys.79.353} electronic Raman scattering,\cite{RevModPhys.83.705} nuclear magnetic
resonance,\cite{RevModPhys.79.175} or resonant inelastic X-ray
scattering.\cite{RevModPhys.73.203} In this regime, electronic correlations tend to be long ranged, non-perturbative, and
strongly momentum dependent.\cite{PhysRevX.5.041041}

The dynamical mean field approximation (DMFT),\cite{PhysRevLett.62.324,PhysRevLett.69.168,PhysRevLett.69.1240,RevModPhys.68.13} originally
developed as an exact theory for the Hubbard model in the infinite dimensional limit, has evolved into a popular approximation for
simulating strongly correlated lattice models. It is based on the realization that, if correlations and interactions are approximated as
local in space, the numerically intractable lattice system can be mapped onto an effective impurity embedded in a self-consistently
determined medium. The impurity problem can then be solved efficiently using numerical impurity solver algorithms.\cite{RevModPhys.83.349}
However, the approximation of a local self-energy is often severe, and momentum dependent effects such as the pseudogap state or d-wave
superconductivity in the Hubbard model\cite{PhysRevLett.110.216405} are missed entirely.

Two types of extensions to address these shortcomings have been proposed. On one hand, cluster methods\cite{RevModPhys.77.1027} (formulated
in real space, such as the cellular dynamical mean field theory\cite{PhysRevB.62.R9283,PhysRevLett.87.186401}, or in k-space, such as the
dynamical cluster approximation\cite{PhysRevB.58.R7475,PhysRevB.61.12739,PhysRevB.88.115101} (DCA)) systematically enlarge the size of the
impurity or `cluster' and recover the exact solution in the infinite cluster size
limit.\cite{PhysRevLett.106.030401,PhysRevB.88.155108,PhysRevX.5.041041,PhysRevLett.95.237001,PhysRevB.72.060411} These methods are
particularly well adapted to study parameter regimes with strong short-ranged correlations but generally miss correlations on length scales
beyond the cluster size. On the other hand, methods that combine perturbative diagrammatic expansions with a non-perturbative solution of a
confined problem,~\cite{2017arXiv170500024R} such as the Dual Fermion (DF) method,\cite{PhysRevB.77.033101} the dynamical vertex
approximation,\cite{PhysRevB.75.045118} trilex,\cite{PhysRevB.92.115109} quadrilex\cite{PhysRevB.94.075159} or -- in the case of non-local
interactions -- the self-energy embedding theory,\cite{PhysRevB.91.121111,1367-2630-19-2-023047} dual bosons,\cite{RUBTSOV20121320} or
EDMFT+GW,\cite{PhysRevB.66.085120,PhysRevLett.90.086402} typically capture some correlations on all length scales but neglect diagrams that
may be important in a strongly correlated regime. While several of these extensions can in principle be made exact by varying a control
parameter (such as cluster size or the perturbation expansion order), the convergence to the exact limit deep in the correlated regime is
often too slow to be practical.

The combination of cluster methods with diagrammatic extensions is therefore desirable, as the exact treatment of short-range correlation in
combination with a diagrammatic treatment of long-range correlations promises a faster convergence to the thermodynamic limit.
Two such methods have been presented before: the combination of the DF method with real-space cluster dynamical mean field
theory\cite{Hafermann2008} and exact diagonalization of isolated clusters\cite{0295-5075-85-2-27007}; and a combination of the k-space
dynamical cluster approximation with the DF method.\cite{PhysRevB.84.155106} 

In this paper we show that the method
presented in Ref.~\onlinecite{PhysRevB.84.155106} contains an additional coarse-graining approximation and becomes exact only in
the limit of cluster size $N_c\rightarrow\infty$ in which the underlying DCA cluster method converges itself and dual fermion
corrections disappear, whereas the methods of Refs.~\onlinecite{Hafermann2008,0295-5075-85-2-27007} become prohibitively expensive for large
clusters. We then present a momentum-space dual fermion cluster method which utilizes an \textit{exact} mapping to the dual fermion
variables, avoiding the coarse graining approximation. We show that the coarse graining approximation used in
Ref.~\onlinecite{PhysRevB.84.155106} is the dominant approximation in practice, and that it can be avoided by breaking translational
symmetry in the solution of the dual fermion problem, leading to a method that is exact for any cluster size.

Our results for self-energies and double occupancies of the Hubbard model obtained on four-site clusters show that there is a substantial
improvement of dual fermion cluster results over single site DF results, both in single-particle and in two-particle quantities. We also
demonstrate that basing the method on a translationally invariant impurity problem leads to a manageable scaling of the impurity vertex
functions, offering the possibility to perform simulations on clusters large enough to exhibit a clear pseudogap and superconducting state.

The remainder of this paper is organized  as follows: In Sec.~\ref{sec:meth} we introduce the the model, derive the dynamical cluster dual
fermion method of Ref.~\onlinecite{PhysRevB.84.155106}, and show how the coarse graining approximation can be avoided by breaking
translational invariance. In Sec.~\ref{sec:res} we show results from our method and compare them to various other methods, including finite
cluster DCA and DF; We conclude and summarize in Sec.~\ref{sec:conc}.

\section{Method}\label{sec:meth}
In this paper, we limit ourselves to studying the Hubbard model on a two-dimensional square lattice with nearest neighbor hopping $t$, chemical potential
$\mu$, and on-site interaction $U$,
\begin{align}
H = \sum_{\langle \li\lj\rangle\sigma} 
t_{ij}
c^{\dagger}_{\li\sigma}c_{\lj\sigma} +U\sum_{\li} 
n_{\li\uparrow}
n_{\li\downarrow}
- \mu\sum_{\li\sigma} c^{\dagger}_{\li\sigma}c_{\li\sigma}.
\label{Eq:Ham}
\end{align}
The operators $c_i^{(\dagger)}$ annihilate (create) a particle at lattice site $\li$, and $n_{\li}=c^{\dagger}_{\li}c_{\li}$ denotes
the particle number at site $\li$.

To start the derivation of cluster methods, we approximate the infinite lattice problem by a finite (but large) number of sites $N_{tot},$
which we divide into $V$ clusters of size $N_c$ each, such that $N_{tot}  = V N_c$. We will refer to the lattice formed by the $V$ clusters
as the `super-lattice'. The position $i$ of a coordinate in the original lattice is then given as the pair of $\cI,\ti$, where
$\ti=\tilde{\mathbf{x}}_1,\cdots,\tilde{\mathbf{x}}_V$ denotes the position of a cluster in the lattice and
$\cI=\mathbf{X}_1,\cdots,\mathbf{X}_{N_c}$ labels sites within each cluster, such that
$\li=\mathbf{\tilde{i}}+\mathbf{I}$.\cite{RevModPhys.77.1027} Eq.~\ref{Eq:Ham} then reads:

\begin{align}
H &= \sum\limits_{\langle\cI\ti,\cJ\tj\rangle\sigma} 
t_{\cI+\ti,\cJ+\tj}
c^{\dagger}_{\cI+\ti\sigma}c_{\cJ+\tj\sigma} + \nonumber \\ +
U\sum\limits_{\cI\ti} &
c^{\dagger}_{\cI+\ti\uparrow}
c_{\cI+\ti\uparrow}
c^{\dagger}_{\cI+\ti\downarrow}
c_{\cI+\ti\downarrow}
-\mu \sum\limits_{\cI\ti\sigma} c^{\dagger}_{\cI+\ti\sigma}c_{\cI+\ti\sigma}.
\label{Eq:HamRS}
\end{align}
The corresponding reciprocal $k$-space is split into $N_c$ cells labeled by momentum $\cK$ with $V$ momenta $\tk$ inside each cell. 
Operators in reciprocal space and in real space are related by the Fourier transforms
\begingroup
\allowdisplaybreaks
\begin{subequations}
\begin{align}
c_{\cI + \ti,\sigma} = \frac{1}{\sqrt{VN_c}} \sum_{\cK,\tk} c_{\cK+\tk,\sigma} e^{i(\cK+\tk)(\cI+\ti)}; \\
c^{\dagger}_{\cI + \ti,\sigma} = \frac{1}{\sqrt{VN_c}} \sum_{\cK,\tk} c^{\dagger}_{\cK+\tk,\sigma} e^{-i(\cK+\tk)(\cI+\ti)}; \\
c_{\cK + \tk,\sigma} = \frac{1}{\sqrt{VN_c}} \sum_{\cI,\ti} c_{\cI+\ti,\sigma} e^{-i(\cK+\tk)(\cI+\ti)}; \\
c^{\dagger}_{\cK + \tk,\sigma} = \frac{1}{\sqrt{VN_c}} \sum_{\cI,\ti} c^{\dagger}_{\cI+\ti,\sigma} e^{i(\cK+\tk)(\cI+\ti)}.
\end{align}
\label{OpFT}
\end{subequations}
\endgroup
Fourier transforming the Hamiltonian of Eq.~\ref{Eq:HamRS} to reciprocal $k$-space according to Eq.~\ref{OpFT} results in
\begin{align}
H = \sum\limits_{\cK\tk\sigma} &\epsilon_{\cK+\tk} c^{\dagger}_{\cK+\tk,\sigma}c_{\cK+\tk,\sigma}
- \mu\sum\limits_{\cK\tk\sigma} c^{\dagger}_{\cK+\tk,\sigma}c_{\cK+\tk,\sigma} + \nonumber \\ 
+ \frac{U}{N_c^2 V^2}&\sum\limits_{\substack{\cK_1\cK_2\cK_3\cK_4 \\\tk_1\tk_2\tk_3\tk_4}} 
c^{\dagger}_{\cK_1+\tk_1\uparrow}
c_{\cK_2+\tk_2\uparrow}
c^{\dagger}_{\cK_3+\tk_3\downarrow}
c_{\cK_4+\tk_4\downarrow} \Lambda,
\label{HubbardModel}
\end{align}
where $\epsilon_{\cK+\tk} = \sum\limits_{\cI\ti,\cJ\tj}
t_{\cI+\ti,\cJ+\tj} e^{i(\cK+\tk)(\cI+\ti - \cJ -\tj)}$ is the lattice dispersion and
\begin{align}
\label{Eq:Laue}
\Lambda =
\sum\limits_{\cI,\ti} e^{-i(\cK_1 + \tk_1 + \cK_3 + \tk_3 - \cK_2 - \tk_2 - \cK_4 - \tk_4)(\cI + \ti)}
\end{align} is
the ``Laue function''\cite{RevModPhys.77.1027} which stems from the Fourier transform defined in Eq.~\ref{OpFT} and guarantees momentum
conservation of the interaction.

For the derivation of the DF formalism (presented in Sec.~\ref{sec:dfderiv}) it is convenient to switch to an action formulation and
Grassmann variables. The action corresponding to Eq.~\ref{HubbardModel} can be expressed as
\begin{align}
S &=S^{(int)} +&\nonumber\\
 +\sum\limits_{\tk} \int d\tau &\sum_{\cK\sigma}
c^{*}_{\cK+\tk,\sigma}(\tau)\left((\partial_{\tau} - \mu) + \epsilon_{\cK+\tk}\right)
c_{\cK+\tk,\sigma}(\tau),
\label{HubbardAction}
\end{align}
where $c_{\cK+\tk,\sigma}\left(\tau\right)$ is a fermionic Grassmann variable, and $S^{(int)}$ is the interacting part of the Hubbard
action, defined as
\begin{align}
S^{(int)} &= \frac{U}{N_c^2V^2}\sum_{\tk_1\tk_2\tk_3\tk_4} \int d\tau \nonumber \\ &
\sum_{\cK_1\cK_2\cK_3\cK_4}
c^{*}_{\cK_1+\tk_1\uparrow}(\tau)
c_{\cK_2+\tk_2\uparrow}(\tau) \times \nonumber \\ &\times
c^{*}_{\cK_3+\tk_3\downarrow}(\tau)
c_{\cK_4+\tk_4\downarrow}(\tau)\Lambda.
\label{HubbardInteraction}
\end{align}

Next, we employ a Fourier transform from imaginary time $\tau$ to Matsubara frequency $w_n=(2n+1)\pi T$ , with
\begin{subequations}
\begin{align}
c_{\omega_n,\cK + \tk,\sigma} &= \frac{1}{\sqrt{\beta}} \int c_{\cK+\tk,\sigma}(\tau) e^{-i\tau\omega_n} d\tau; \\
c^{*}_{\omega_n,\cK + \tk,\sigma} &= \frac{1}{\sqrt{\beta}} \int c^{*}_{\cK+\tk,\sigma}(\tau) e^{i\tau\omega_n}d\tau.
\end{align}
\end{subequations}
The resulting Hubbard model action in reciprocal space and Matsubara frequencies then reads
\begin{widetext}
\begin{align}
S = - \sum_{n\cK\tk\sigma}c^{*}_{\omega_n,\cK+\tk,\sigma}(i\omega_n &+ \mu - \epsilon_{\cK+\tk})c_{\omega_n,\cK+\tk,\sigma} + \nonumber
\\ 
+&\frac{U}{N_c^2V^2\beta}\sum_{\substack{n,m,l \\ \cK_1\cK_2\cK_3\cK_4\\ \tk_1\tk_2\tk_3\tk_4}}
c^{*}_{\omega_n,\cK_1+\tk_1,\uparrow}
c_{\omega_n+\Omega_l,\cK_2+\tk_2,\uparrow}
c^{*}_{\omega_m+\Omega_l,\cK_3+\tk_3,\downarrow}
c_{\omega_m,\cK_4+\tk_4,\downarrow} \Lambda,
\label{HubbardActionM}
 \end{align}
\end{widetext}
with bosonic transfer frequency $\Omega_n=\frac{2n\pi}{\beta}$.
Eq.~\ref{HubbardActionM} is the starting point for the cluster DF derivations presented in following sections.

\subsection{Dual Fermion method}\label{sec:dfderiv}
In this section, we give a short overview of the single site  DF method.\cite{PhysRevB.77.033101} A detailed discussion of the formalism can
be found in Ref.~\onlinecite{PhysRevB.79.045133}. The DF method aims to use DMFT as a starting point and reintroduce all non-local correlation
effects via a diagrammatic perturbation series around DMFT.

The construction of the DF approximation is typically done in two steps. First, a DMFT impurity model that describes local correlations is
constructed and solved self-consistently using numerical impurity solvers.\cite{RevModPhys.83.349} Second, at the cost of the introduction
of new fermionic operators, a perturbation theory of the original lattice model in terms of local quantities is formulated.
In this construction, the DMFT  is a zeroth-order approximation, and non-local correlations are obtained by a diagrammatic perturbation
series of the DF problem. The DF method utilizes a formally exact mapping from the original lattice problem to so-called `dual' variables,
and is formulated in terms of a hierarchy of (two-particle, three-particle, etc) impurity vertices and modified `dual' propagators.  In
practical implementation of the DF method, the hierarchy of impurity vertices is then usually truncated to the two-particle vertex, and only
second order or ladder-type diagrams of the perturbation series are considered.

The dual fermion method has provided important insights to electron correlation effects in the Hubbard
model\cite{PhysRevB.79.045133,PhysRevLett.102.206401,PhysRevB.92.144409,PhysRevB.90.235132} and in disordered lattice
systems.\cite{PhysRevB.87.134208} It has also found applications in the study of the Falicov-Kimball
model,\cite{PhysRevLett.112.226401,PhysRevB.95.155130,PhysRevB.93.195105} where the equations can be solved exactly.
Unfortunately, strong non-local fluctuations, such as those present near half-filling in the two-dimensional Hubbard model, render the local
DMFT approximation less suitable as a starting point for DF perturbation expansion.
For this reason, starting points that include non-local correlations\cite{Hafermann2008} are needed to get an accurate
representation of the model system at low DF expansion order. Developing such a starting point is the topic of the present paper.

\subsection{DCA method and DCA approximation}\label{sec:DCA}
Both Ref.~\onlinecite{PhysRevB.84.155106} and our momentum-space cluster DF method employ the DCA as a zeroth order approximation around
which the DF perturbation expansion is performed. In this section, we briefly review dynamical cluster approximation itself. While we only
outlined the main ideas used in the DCA construction, a more detailed description of the DCA method can be found in
Ref.~\onlinecite{RevModPhys.77.1027}.

DCA is based on an approximation of the infinite, intractable lattice model by a periodic cluster of size $N_c$  embedded in a
self-consistently determined host defined by hybridization function $\Delta_{\omega_n\cK}$. Conceptually, this is done by partitioning the
lattice with $N$ sites into clusters containing $N_c$ sites which, in reciprocal space, is equivalent to the division of the Brillouin zone
of the underlying lattice into clusters with $N_c$ cells centered at the reciprocal sub-lattice vectors $\cK$. The clusters are subject to
periodic boundary conditions.

In its construction, the  DCA employs two approximations. First, DCA assumes that the self-energy of the original lattice and of the cluster
are related by $\Sigma(\lk,\omega) \approx \Sigma_c(\cK,\omega)$, where $\Sigma(\lk,\omega)$ denotes the lattice self-energy and
$\Sigma_c(\cK,\omega)$ the cluster self-energy, and $\lk=\cK+\tk$. In other words, the self-energy is assumed to be constant within each
cluster `cell'. Second, DCA assumes that each cluster is indistinguishable in the super-lattice and that momentum is only conserved within
the cluster. This is equivalent to the approximation of momentum conservation at each vertex by neglecting the phase factors $e^{i\tk\ti}$
and $e^{i\tk\cI}$, respectively, in Eq.~\ref{Eq:Laue}.
This second approximation, sometimes called the `coarse-graining' of the interaction term, results in replacing the Hubbard model Laue
function (ensuring momentum conservation on a lattice) with the DCA Laue function
\begin{align}\label{eq:DCALaue}
\Lambda \approx \Lambda_{DCA} = \sum\limits_{\cI} e^{-i(\cK_1 + \cK_3 - \cK_2 -\cK_4)\cI}
\end{align}
(which ensures momentum conservation within a cluster).\cite{RevModPhys.77.1027}\textsuperscript{,}
\footnote{Note that in Eq.~\ref{Eq:Laue}, for any $\alpha$ component of cluster momenta $\cK_{\alpha} = \frac{2\pi l_{\alpha}}{a}$ and the corresponding component of the
super-lattice real-space vector $\ti_{\alpha} = a m_{\alpha}$, where $a$ is the super-lattice constant and $m_{\alpha}$ and $l_{\alpha}$ are
integer, the phase factors $e^{i\cK\ti}$ are always $1$.}

The standard DCA derivation then approximates Eq.~\ref{HubbardActionM} by a cluster action of the form
\cite{RevModPhys.77.1027}
\begin{align}
S_{c} =- \sum_{n\cK\sigma} &{c}^{*}_{\omega_n,\cK,\sigma} G^{0}_{c}(\omega_n)^{-1}{c}_{\omega_n,\cK,\sigma} + S^{(c)}_{int}.
\label{ClusterAction}
\end{align}
In this equation, $S^{(c)}_{int}$ is the interacting part of the cluster action defined as 
\begin{align}
S^{(c)}_{int} =&
\frac{U}{N_c^2\beta}\sum_{\substack{n,m,l \\ 
\cK_1\cK_2\cK_3\cK_4}}  {c}^{*}_{\omega_n,\cK_1,\uparrow} {c}_{\omega_n+\Omega_l,\cK_2,\uparrow} \times \nonumber \\ 
&\times {c}^{*}_{\omega_m+\Omega_l,\cK_3,\downarrow} {c}_{\omega_m,\cK_4,\downarrow}\Lambda_{DCA},
\label{ClusterInteraction}
\end{align}
and $G_c^{0}(\bf K, \omega_n) = (i\omega_n+\mu -
\bar{\epsilon}_{\cK} - \Delta_{\omega_n\cK})^{-1}$ is the bare cluster Green's function, with level energies $\bar{\epsilon}_{\cK}$ and
hybridization function $\Delta_{\omega_n\cK}$ chosen such that $\Sigma(\lk)=\Sigma_c(\cK)$.\cite{RevModPhys.77.1027} A solution is typically
obtained iteratively, starting from some initial guess for $\Delta_{\omega_n\cK}$, until the convergence criterion
$\Sigma(\lk)=\Sigma_c(\cK)$ is satisfied.

\subsection{DCA dual fermion methods}
\label{subsec:meth}
In this section, we derive two cluster dual fermion methods that use the dynamical cluster approximation as a starting point. In
Sec.~\ref{subsec:meth1} we show that, if the Laue function approximations of Sec.~\ref{sec:DCA} is kept, the approximate method of
Ref.~\onlinecite{PhysRevB.84.155106} is obtained. Subsequently, in Sec.~\ref{subsec:meth2}, we show how our momentum-space cluster DF method
avoids this approximation.

DCA dual fermion methods aim to perform a dual fermion expansion around cluster problems solved by the dynamical cluster approximation.
To this end, Eq.~\ref{HubbardActionM} is expressed in terms of DCA cluster actions, Eq.~\ref{ClusterAction}, with additional inter-cluster
correction terms to systematically recover the approximations introduced within the DCA framework.

\subsubsection{DCA dual fermions with Laue approximation}
\label{subsec:meth1}
As we have shown in Sec.~\ref{sec:DCA}, in addition to restricting the self-energy dependence to the cluster momenta $\Sigma(\cK,\omega_n)$ only,
the DCA utilizes the coarse-graining of the interaction term approximation, resulting in neglecting of $e^{i\tk\ti}$ and $e^{i\tk\cI}$ phase
factors in the lattice Laue function. In this section, we show a method that will allow the construction of a fully momentum dependent self
energy  $\Sigma(\lk,\omega_n)$. However, the DF mapping used will remain approximate. In particular, we will include the phase factor
$e^{i\tk\ti}$, but keep the approximation for the phase factor $e^{i\tk\cI}$ to have cluster momentum conservation inside the cluster.

To start the construction, we rewrite the lattice action, Eq.~\ref{HubbardActionM}, in terms of a sum of DCA cluster actions with additional
non-local correction terms. To make the clusters translationally invariant, we set $e^{i\tk\cI}=1$ in the lattice Laue function,
Eq.~\ref{Eq:Laue}, resulting in
\begin{align}
\Lambda \approx& \sum\limits_{\cI,\ti} e^{-i(\cK_1 + \cK_3 - \cK_2 -\cK_4)\cI}e^{-i(\tk_1 + \tk_3 - \tk_2 -\tk_4)\ti} =\nonumber \\
=&\Lambda_{DCA} \sum_{\ti}e^{-i(\tk_1 + \tk_3 - \tk_2 -\tk_4)\ti}.
\label{LaueApproximation}
\end{align}
This allows us to define the annihilation operator at (real-space) cluster position $\ti$ and (cluster reciprocal coordinate) $\cK$ using
Eq.~\ref{OpFT} to approximate the interacting part of the lattice action while keeping the kinetic part exact
\begin{align}
c_{\cK,\ti,\sigma} = \frac{1}{\sqrt{N_c}} \sum_{\cI} c_{\cI+\ti,\sigma} e^{-i\cK\cI},
\label{approxFT}
\end{align}
and to rewrite the lattice action of Eq.~\ref{HubbardActionM} as
\begin{align}
S &= - \sum_{n\cK\tk\sigma}c^{*}_{\omega_n,\cK+\tk,\sigma}(i\omega_n + \mu - \epsilon_{\cK+\tk})c_{\omega_n,\cK+\tk,\sigma} + \nonumber \\+&
\sum_{\ti} S^{(\ti)}_{int},
\label{HubbardActionDCAL}
 \end{align}
where $S^{(\ti)}_{int}$ is the interacting part of the action 
\begin{align}
S^{(\ti)}_{int} &= \frac{U}{N_c^2\beta}\sum_{\substack{n,m,l \\ 
\cK_1\cK_2\cK_3\cK_4}}  {c}^{*}_{\omega_n,\cK_1,\ti,\uparrow} {c}_{\omega_n+\Omega_l,\cK_2,\ti,\uparrow} \times \nonumber \\ 
&\times {c}^{*}_{\omega_m+\Omega_l,\cK_3,\ti,\downarrow} {c}_{\omega_m,\cK_4,\ti,\downarrow}\Lambda_{DCA},
\label{HubbardActionDCALInt}
 \end{align}
which is equivalent to the definition of the interacting part of the DCA effective action of. Eq.~\ref{ClusterInteraction} for each
cluster at position $\ti$ in the super-lattice.

Following the DF construction, now we can map approximated lattice action, Eq.~\ref{HubbardActionDCAL}, to a set of DCA cluster actions,
Eq.~\ref{ClusterAction}, with additional inter-cluster corrections. For this we add and subtract 
$\sum\limits_{n\cK\ti\sigma} c^{*}_{\omega_{n},\cK,\ti,\sigma} (\bar{\epsilon}_{\cK} + \Delta_{\omega_n\cK}) c_{\omega_{n},\cK,\ti,\sigma}$
into the lattice action of Eq.~\ref{HubbardActionDCAL},
\begin{align}
S = -& \sum_{n\cK\tk\sigma} c^{*}_{\omega_n,\cK+\tk,\sigma}\left(i\omega_n + \mu - \epsilon_{\cK+\tk}\right)
c_{\omega_n,\cK+\tk,\sigma} + \nonumber \\ +&\sum_{\ti} S^{(\ti)}_{int} +
\sum_{n\cK\ti\sigma}c^{*}_{\omega_n,\cK,\ti,\sigma}(\bar{\epsilon}_{\cK} + \Delta_{\omega_n\cK})c_{\omega_n,\cK,\ti,\sigma} - \nonumber \\
-&\sum_{n\cK\ti\sigma}c^{*}_{\omega_n,\cK,\ti,\sigma}(\bar{\epsilon}_{\cK} +\Delta_{\omega_n\cK}) c_{\omega_n,\cK,\ti,\sigma}.
\end{align}
Since the cluster patch density obeys $\sum\limits_{\ti}c^{*}_{\omega_n,\cK,\ti,\sigma}c_{\omega_n,\cK,\ti,\sigma} = \sum\limits_{\tk}
c^{*}_{\omega_n,\cK+\tk,\sigma}c_{\omega_n,\cK+\tk,\sigma}$, we can transform the last term in the super-lattice reciprocal space:
\begin{align}
S =-&\sum_{n\cK\ti\sigma} c^{*}_{\omega_n,\cK,\ti,\sigma}\left(i\omega_n + \mu -\bar{\epsilon}_{\cK} -
\Delta_{\omega_n\cK}\right)c_{\omega_n,\cK,\ti,\sigma} + \nonumber \\ &+
\sum_{\ti} S^{(\ti)}_{int}
-\sum_{n\cK\tk\sigma}c^{*}_{\omega_{n},\cK+\tk,\sigma}(\Delta_{\omega_n\cK} +  \nonumber \\ &+
\bar{\epsilon}_{\cK} -\epsilon_{\cK+\tk})c_{\omega_{n},\cK+\tk,\sigma}.
\end{align}
Or
\begin{align}
S &=  \sum_{\ti} S^{(\ti)}_{c} \\ \nonumber
 -\sum_{n\cK\tk\sigma}&c^{*}_{\omega_{n},\cK+\tk,\sigma}(\Delta_{\omega_n\cK} +
\bar{\epsilon}_{\cK} -\epsilon_{\cK+\tk})c_{\omega_{n},\cK+\tk,\sigma}.
\end{align}
So far, we have rewritten our lattice action in terms of the set of cluster actions 
$S^{(\ti)}_c$ plus the last terms, which describe the coupling
between the super-sites. 

Next, we introduce the dual fermion variables $(\xi^*,\xi)$ by performing a Hubbard-Stratonovich transformation in the last term for each
inter-cluster momenta $\tk$, choosing $\alpha = \{\omega_n,\cK,\tk\}$, $\beta=\{\omega_n,\cK',\tk\}$, $a_{\alpha\beta} =
[(\Delta_{\omega_n\cK} + \bar{\epsilon}_{\cK} -\epsilon_{\cK+\tk})\delta_{\cK\cK'}]^{-1}$ and $b$ as an arbitrary function of $\omega_n$, $\cK$ and $\sigma$:
\begin{align}
e^{c^{*}_{\alpha}a^{-1}_{\alpha\beta}c_{\beta}} = \frac{1}{\det(b_{\alpha}a_{\alpha\beta}b_{\beta})} \int& \nonumber \\ 
 e^{-\xi^{*}_{\alpha} b_{\alpha}a_{\alpha\beta}b_{\beta} \xi_{\beta} + \xi^{*}_{\alpha} b_{\alpha} c_{\alpha} + c^{*}_{\alpha}
 b_{\alpha} \xi_{\alpha}}& \mathcal{D}[\xi^{*},\xi],
\label{HubbardStratonovich}
\end{align}
such that
\begin{align}\label{Eq:Sdual}
S &= \sum_{\ti} S^{(\ti)}_{c} +  \sum_{K\tk} \xi^{*}_{K\tk} b_{K}\frac{1}{(\Delta_{\omega_n\cK} +
\bar{\epsilon}_{\cK} -\epsilon_{\cK+\tk})}b_{K} \xi_{K\tk}
- \nonumber \\ -&
\sum_{K\tk} (\xi^{*}_{K\tk} b_{K} c_{K\tk} + c^{*}_{K\tk} b_{K} \xi_{K\tk}).
\end{align}
Here $K=\{\omega_n\cK\sigma\}$ is composite index for shorthand notation, and we choose $b(K) = - {G^{(c)}_K}^{-1}$. Since $b(K)$ is
independent of $\tilde{k}$ we can Fourier transform the last term in above Eq. to the super-lattice real space, i.e. 
$\sum_{K\tk} (\xi^{*}_{K\tk} b_{K} c_{K\tk} + c^{*}_{K\tk} b_{K} \xi_{K\tk})= \sum_{K\ti} (\xi^{*}_{K\ti}
b_{K} c_{K\ti} + c^{*}_{K\ti} b_{K} \xi_{K\ti})$. Eq.~\ref{Eq:Sdual} then reads
\begin{align}
S &= 
\sum_{\ti} S^{(\ti)}_{c} - \sum_{K\ti} (\xi^{*}_{\ti K} b_{K} c_{K\ti} + c^{*}_{K\ti} b_{K} \xi_{K\ti}) +\nonumber \\
&+ \sum_{K\tk} \xi^{*}_{K\tk} b_{K}\frac{1}{(\Delta_{\omega_n\cK} +
\bar{\epsilon}_{\cK} -\epsilon_{\cK+\tk})}b_{K} \xi_{K\tk}.
\end{align}

This allows us to complete the mapping to the DF lattice by integrating out the lattice fermionic Grassmann variables $c^*_{K \ti},c_{K\ti}$
for each super-lattice site $\ti$ separately:
\begin{align}
\int &e^{-S^{(\ti)}_{c}} e^{-\sum\limits_{K} \left(\xi^{*}_{K\ti} {G^{(c)}_K}^{-1} c_{K\ti} + c^{*}_{K\ti} {G^{(c)}_K}^{-1}
\xi_{K\ti}\right)} \mathcal{D}\left[c^{*}_{\ti},c_{\ti}\right] = \nonumber \\
=&\mathcal{Z}_{c} \left\langle e^{-\sum\limits_{K} \left(\xi^{*}_{K\ti} {G^{(c)}_K}^{-1} c_{K\ti} + c^{*}_{K\ti} {G^{(c)}_K}^{-1}
\xi_{K\ti}\right)} \right\rangle_{c},
\label{IntegrationOut1}
\end{align}
where $\langle\ldots\rangle_{c}$ corresponds to averaging in terms of DCA cluster action. The original lattice action problem can then be
expressed using only dual operators and cluster observables:
\begin{align}
S\left[ \xi^{*},\xi\right] &= \sum_{n,\cK+\tk,\sigma} \xi^{*}_{\omega_{n},\cK+\tk,\sigma}
{G^{(c)}_{K}}^{-1} \times \nonumber \\ \times&
((\Delta_{\omega_n\cK} + \bar{\epsilon}_{\cK} -\epsilon_{\cK+\tk}))^{-1} {G^{(c)}_{K}}^{-1}
\xi_{\omega_{n},\cK+\tk,\sigma} + \nonumber \\ 
+  \sum\limits_{\ti}&\left(\sum\limits_{n\cK\sigma}\xi^{*}_{\omega_{n},\cK,\ti,\sigma} {G^{(c)}_{K}}^{-1}  \xi_{\omega_{n},\cK,\ti,\sigma} 
+ V\left[\xi^{*}_{\ti},\xi_{\ti}\right]\right) = \nonumber \\
= - \sum_{n\cK\tk\sigma} &\xi^{*}_{\omega_{n},\cK+\tk,\sigma} \tilde{G}^{(0)}_{\omega_{n}\cK\tk\sigma}{}^{-1}\xi_{\omega_{n},\cK+\tk,\sigma}
+ \sum\limits_{\ti}V\left[\xi^{*}_{\ti},\xi_{\ti}\right].
\end{align}
Here $\tilde{G}^{(0)}_{\omega_{n}\cK\tk\sigma} = -\frac{G^{(c)}_{K}G^{(c)}_{K}}{ G^{(c)}_{K} + (\Delta_{\omega_n\cK} + \bar{\epsilon}_{\cK}
-\epsilon_{\cK+\tk})^{-1}}$ is a bare dual Green's function, and $V[\xi^{*}_{\ti},\xi_{\ti}]$ is the dual interaction term
\begin{align}
V\left[\xi^{*}_{\ti},\xi_{\ti}\right] = \sum_{n=2}^{\infty}\frac{(-1)^{n-1}}{(n!)^2} &\sum_{K_1\ldots K_{2n}} \gamma^{(2n)}_{c,K_1\ldots
K_{2n}} \times \nonumber \\ \times
[\prod_{m=0}^{n-1} \xi^{*}_{\ti,K_{m n + 1}}\xi_{\ti,K_{m n + 2}}]& \delta_{\sum_{i\in odd}K_i,\sum_{j\in even}K_j} .
\label{eq:dual_potential}
\end{align}
Here $\gamma^{(2n)}_c$ is the antisymmetrized $n-$particle reducible cluster vertex function. This derivation matches the previously published
results.\cite{PhysRevB.84.155106} Note, however, that it intrinsically relies on the approximation to $\Lambda$,
Eq.~\ref{LaueApproximation}. While this approximation becomes exact in the limit of $N_c\rightarrow \infty$ (where DCA becomes exact and all
dual fermion corrections disappear), the method is approximate for any finite cluster size. This is very different from single site dual
fermions or the real-space cluster methods, which recover the exact result provided that all dual corrections are summed. In
Sec.~\ref{sec:res}, we will show that -- at least for the parameter regime studied -- the effect of this Laue approximation is considerable.

\subsubsection{Momentum-Space Cluster Dual Fermion Method}
\label{subsec:meth2}
In this section we present a cluster DF method which avoids  the approximation to the Laue function. As a consequence, the dual fermion
mapping will be exact for any cluster size, meaning that if all dual fermion diagrams to all orders are summed, the exact result will be
recovered independent of the quality of the underlying DCA approximation.

To preserve the lattice momentum conservation and express the lattice action of Eq.~\ref{HubbardInteraction} in terms of DCA cluster
actions, we separately define Fourier transforms for intra-($I$) and inter ($\ti$)-cluster momenta:
\begingroup
\allowdisplaybreaks
\begin{subequations}
\begin{align}
c_{\cI,\ti,\sigma} &= \frac{1}{\sqrt{V}} \sum_{\tk} c_{\cI,\tk,\sigma} e^{i\tk\ti} ;\\ 
c_{\cI,\ti,\sigma} &= \frac{1}{\sqrt{N_c}} \sum_{\cK} c_{\cK,\ti,\sigma} e^{i\cK\cI};\\
c_{\cK,\tk,\sigma} &= \frac{1}{\sqrt{V}} \sum_{\ti} c_{\cK,\ti,\sigma} e^{-i\tk\ti}; \\
c_{\cK,\tk,\sigma} &= \frac{1}{\sqrt{N_c}} \sum_{\cI} c_{\cI,\tk,\sigma} e^{-i\cK\cI}.
\end{align}
\end{subequations}
\endgroup
In this case the lattice action is
\begin{widetext}
\begin{align}
S &= - \sum\limits_{n,\tk} \sum_{\cK\cK'\sigma}
c^{*}_{\omega_n,\cK,\tk,\sigma}\left((i\omega_n + \mu)_{{\delta_{\cK\cK'}}} - \hat{\epsilon}_{\cK\cK'\tk}\right)
c_{\omega_n,\cK',\tk,\sigma} + \nonumber \\
+&\frac{U}{N_c V\beta}\sum_{\substack{n,m,l\\\tk\tk'\tq}} \sum_{\cK\cK'\cQ}
c^{*}_{\omega_n,\cK,\tk,\uparrow}
c_{\omega_n+\Omega_l,\cK+\cQ,\tk+\tq,\uparrow}
c^{*}_{\omega_m+\Omega_l,\cK'+\cQ,\tk'+\tq,\downarrow}
c_{\omega_m,\cK',\tk',\downarrow}.
\label{HubbardAction2}
\end{align}
\end{widetext}

Here $\hat{\epsilon}_{\cK\cK'\tk}$ is the reciprocal space representation of the hopping term. This hopping term becomes non-diagonal in
reciprocal cluster vectors $K$ and $K'$ due to the breaking of the translational invariance.
\begin{align}\label{Eq:epsilonhat}
\hat{\epsilon}_{\cK\cK'\tk} = \frac{1}{N_c} \sum_{\cI\cJ\ti\tj} t_{\cI+\ti,\cJ+\tj} e^{i(\cK\cI - \cK'\cJ + \tk(\ti - \tj))}.  
\end{align}

In order to express our lattice action in terms of the translationally invariant DCA cluster action, first we need to express interaction
term in super-lattice real-space as
\begin{align}
\frac{U}{N_c V\beta}\sum_{\substack{n,m,l\\\tk\tk'\tq}} \sum_{\cK\cK'\cQ} &
c^{*}_{\omega_n,\cK,\tk,\uparrow}
c_{\omega_n+\Omega_l,\cK+\cQ,\tk+\tq,\uparrow} \times \nonumber \\ \times&
c^{*}_{\omega_m+\Omega_l,\cK'+\cQ,\tk'+\tq,\downarrow}
c_{\omega_m,\cK',\tk',\downarrow} = \nonumber \\ =
\frac{U}{N_c\beta}\sum_{\substack{n,m,l\\\ti}} \sum_{\cK\cK'\cQ} &
c^{*}_{\omega_n,\cK,\ti,\uparrow}
c_{\omega_n+\Omega_l,\cK+\cQ,\ti,\uparrow} \times \nonumber \\ \times&
c^{*}_{\omega_m+\Omega_l,\cK'+\cQ,\ti,\downarrow}
c_{\omega_m,\cK',\ti,\downarrow}.
\end{align}

To express Eq.~\ref{HubbardAction2} in terms of the DCA cluster action of Eq.~\ref{ClusterAction}, we now add and subtract
$\sum\limits_{n,\cK\ti\sigma} c^{*}_{\omega_{n},\cK,\ti,\sigma} (\bar{\epsilon}_{\cK} + \Delta_{\omega_{n}\cK})
c_{\omega_{n},\cK,\ti,\sigma}$ into lattice action term:

\begin{align}
S = -& \sum_{\substack{n\\\cK,\cK'\\\tk,\sigma}} c^{*}_{\omega_{n},\cK,\tk,\sigma}\left((i\omega_n + \mu)_{{\delta_{\cK\cK'}}} -
\hat{\epsilon}_{\cK\cK'\tk}\right) c_{\omega_{n},\cK',\tk,\sigma} + \nonumber\\
+& \sum_{\ti} S^{(\ti)}_{int} +
\sum_{n\cK\ti\sigma}c^{*}_{\omega_n,\cK,\ti,\sigma}(\bar{\epsilon}_{\cK} +
\Delta_{\omega_n\cK})c_{\omega_n,\cK,\ti,\sigma} - \nonumber \\
-&\sum_{n\cK\ti\sigma}c^{*}_{\omega_n,\cK,\ti,\sigma}(\bar{\epsilon}_{\cK} +\Delta_{\omega_n\cK})
c_{\omega_n,\cK,\ti,\sigma}.
\end{align}
Since the DCA cluster quantities $\bar{\epsilon}_{\cK}$ and $\Delta_{\omega_n,\cK}$ are independent of cluster position in the super
lattice, we transform the last term into super-lattice reciprocal space $(\tk)$:
\begin{align}
S=-&\sum_{n\cK\ti\sigma} c^{*}_{\omega_n,\cK,\ti,\sigma}\left(i\omega_n + \mu -\bar{\epsilon}_{\cK} - \Delta_{\omega_n\cK}\right)
c_{\omega_n,\cK,\ti,\sigma} +\nonumber \\  +& \sum_{\ti} S^{(\ti)}_{int} - \nonumber \\ 
-&\sum_{n\cK\cK'\tk\sigma}c^{*}_{\omega_{n},\cK,\tk,\sigma}((\Delta_{\omega_n\cK} + \bar{\epsilon}_{\cK})_{\delta_{\cK\cK'}} -\hat{\epsilon}_{\cK\cK'\tk})
c_{\omega_{n},\cK',\tk,\sigma}.
\end{align}
Or, using the Eq.~\ref{ClusterAction} for the cluster action,
\begin{align}
S=&\sum_{\ti} S^{(\ti)}_{c} - \nonumber \\ -&\sum_{n\cK\cK'\tk\sigma}c^{*}_{\omega_{n},\cK,\tk,\sigma}((\Delta_{\omega_n\cK} +
\bar{\epsilon}_{\cK})_{\delta_{\cK\cK'}} -\hat{\epsilon}_{\cK\cK'\tk})c_{\omega_{n},\cK',\tk,\sigma}.
\end{align}

Next, we introduce dual fermions via a Hubbard-Stratonovich transformation, Eq.~\ref{HubbardStratonovich}, choosing
$a_{\alpha\beta} = (\Delta_{\omega_n\cK} + \bar{\epsilon}_{\cK} - \hat{\epsilon}_{\cK\cK'\tk})^{-1}$, and $b$ as an arbitrary function of
$\omega_n$, $\cK$ and $\sigma$. This expresses the lattice action in
terms of the cluster action and dual Grassman variables:
\begin{align}
S &= \sum_{\ti} S^{(\ti)}_{c} + \nonumber \\ &+\sum_{KK'\tk} \xi^{*}_{K,\tk} b_{K}((\Delta_{\omega_n\cK} + \bar{\epsilon}_{\cK})_{\delta_{\cK\cK'}}
-\hat{\epsilon}_{\cK\cK'\tk})^{-1}b_{K'} \xi_{K',\tk} -\nonumber \\ -& \sum_{K\tk} (\xi^{*}_{K,\tk} b_{K} c_{K,\tk} + c^{*}_{K,\tk}
b_{K} \xi_{K,\tk}) = \nonumber \\
&= \sum_{\ti} S^{(\ti)}_{c} - \sum_{K\ti} (\xi^{*}_{K,\ti} b_{K} c_{K,\ti} + c^{*}_{K,\ti} b_{K} \xi_{K,\ti}) + \nonumber \\ +&
 \sum_{KK'\tk} \xi^{*}_{K,\tk} b_{K}((\Delta_{\omega_n\cK} + \bar{\epsilon}_{\cK})_{\delta_{\cK\cK'}} -\hat{\epsilon}_{\cK\cK'\tk})^{-1}b_{K'} \xi_{K',\tk},
\end{align}
where $K$ and $K'$ have been introduced as shorthand notation for $K=\{\cK,\omega_n,\sigma\}$ and $K'=\{\cK',\omega_n,\sigma\}$ respectively. 

Considering the first two terms in the last equation, we can now complete the mapping to the DF lattice by integrating out the lattice
fermion Grassmann variables for each super-lattice site $\ti$ separately and express the original lattice action via dual fermion Grassmann
variables $\xi^{*}$ and $\xi$ only, choosing $b=-{G^{(c)}_{K}}^{-1}$:
\begin{align}
S\left[ \xi^{*},\xi\right] &= \sum_{n\cK\cK'\tk\sigma} \xi^{*}_{\omega_{n},\cK,\tk,\sigma} \times \nonumber \\ \times
{G^{(c)}_{K}}^{-1} &\left[((\Delta_{\omega_n\cK} + \bar{\epsilon}_{\cK})_{\delta_{\cK\cK'}}
-\hat{\epsilon}_{\cK\cK'\tk})\right]^{-1}{G^{(c)}_{K'}}^{-1} \xi_{\omega_{n},\cK',\tk,\sigma} + \nonumber \\
+\sum\limits_{\ti}&\left(\sum\limits_{n\cK\sigma}\xi^{*}_{\omega_{n},\cK,\ti,\sigma} {G^{(c)}_{K}}^{-1} \xi_{\omega_{n},\cK,\ti,\sigma} +  V\left[\xi^{*}_{\ti},\xi_{\ti}\right]\right) =\nonumber \\
=& -\sum_{n\cK\cK'\tk\sigma} \xi^{*}_{\omega_{n},\cK,\tk,\sigma}
\tilde{G}^{(0)}_{\omega_{n}\cK\cK'\tk\sigma}{}^{-1}\xi_{\omega_{n},\cK',\tk,\sigma} + \nonumber \\ &+
\sum\limits_{\ti}V\left[\xi^{*}_{\ti},\xi_{\ti}\right],
\label{DualAFAction}
\end{align}
where $\tilde{G}^{(0)}_{\omega_{n}\cK\cK'\tk\sigma} = -G^{(c)}_{K} [G^{(c)}_{K'} + ((\Delta_{\omega_n\cK} +
\bar{\epsilon}_{\cK})_{\delta_{\cK\cK'}} -\hat{\epsilon}_{\cK\cK'\tk})^{-1}]^{-1} G^{(c)}_{K'}$ is a bare dual Green's function and the dual potential,
$V[\xi^{*}_{\ti},\xi_{\ti}]$, remains the same as in Eq.~\ref{eq:dual_potential}.
Detailed derivations of the dual fermion potential construction are given in the supplement.~\cite{SuppMat}

Once the DF perturbation expansion around the DCA problem is performed, it yields correction to the DCA self-energy. The relation between
the dual self-energy $\tilde{\Sigma}_{\cK\cK'\tk}$ and the lattice self-energy $\Sigma_{\cK\cK'\tk}$ is the (see Supplementary material for
the detailed derivation)~\cite{SuppMat}
\begin{align}
\Sigma_{\cK\cK'\tk} = \Sigma^{(c)}_{\cK}\delta_{\cK,\cK'} + \left(1 +
\Sigma_{\cK\cK'\tk}G^{(c)}_{K'}\right)^{-1}\tilde{\Sigma}_{\cK\cK'\tk},
\label{LatticeSigma}
\end{align}
where $\Sigma^{(c)}_{\cK}$ is the DCA cluster self-energy. 

Since in an exact solution of the dual equations we recover the exact, translationally invariant lattice self-energy, we can now transform
$\Sigma_{\cK\cK'\tk}$ into real space separately for intra- and inter-cluster momenta and then perform a regular Fourier transform to obtain
the lattice self-energy
\begin{align}
\Sigma_{\cI,\cJ,\tk} = \frac{1}{N_c^2} \sum\limits_{\cK\cK'} \Sigma_{\cK\cK'\tk} e^{-i (\cI\cK - \cJ\cK') }; \\
\Sigma_{\lk} = \Sigma_{\cK + \tk} = \sum_{\cI,\cJ} \Sigma_{\cI,\cJ,\tk} e^{i (\cI - \cJ)(\cK+\tk) }.
\label{SymmLatticeSigma}
\end{align}

As is evident from Eq.~\ref{Eq:epsilonhat} and Eq.~\ref{DualAFAction}, the cluster dispersion $\hat{\epsilon}_{\cK\cK'\tk}$ and the dual
quantities $\tilde{G}^{(0)}_{\omega_{n}\cK\cK'\tk\sigma}$ and $\tilde{\Sigma}_{\cK\cK'\tk}$ are not translationally invariant in cluster
momenta. The method can therefore be considered as a hybrid
between the dual fermion expansion of cluster DMFT\cite{Hafermann2008} (where both cluster and DF quantities break translational symmetry)
and the DCA DF method of Ref.~\onlinecite{PhysRevB.84.155106} and Sec.~\ref{subsec:meth1}, where neither does, at the cost of introducing an
additional approximation to momentum conservation.

In practice, an approximate solution of the dual fermion equations (e.g. by only considering low-order or ladder diagrams, or by neglecting
any higher order vertices) will result in offdiagonal elements, which can be used to assess the quality of the dual fermion solution.

Similar to the original DF method and the CDMFT DF expansion of Ref.~\onlinecite{Hafermann2008} but unlike the method of
Ref.~\onlinecite{PhysRevB.84.155106}, our momentum space cluster DF method presented in this section recovers the exact solution if all dual
diagrams are summed up to all orders, irrespective of the quality of the underlying DCA solution or the size of the underlying cluster.
However, if the underlying DCA solution already captures the salient aspects of the solution, dual corrections are expected to be small and
perturbative, such that the dual series converges quickly.

In the following, we briefly describe the numerical procedure we use to perform the momentum-space cluster DF analysis, as well as the
numerical scaling. The calculation of this approximation consists of four parts. First, we obtain the converged DCA solution of the problem.
Second, we extract the DCA vertex from the DCA four-point correlation function; Third, we perform the DF perturbation expansion and obtain
the dual self-energies. Finally, we calculate the lattice self-energies. The numerically expensive part is the second part, where a strongly
correlated many-body problem needs to be solved, typically using quantum Monte Carlo methods, and a vertex function has to be extracted.
Away from certain high-symmetry points, all cluster solvers scale exponentially in system size, making access to large cluster sizes
difficult. Measuring the two-particle vertex functions for enough frequencies that reliable dual fermion calculations can be performed, and for all
momenta triples in the cluster, dominates the time to solution. Reducing the scaling from $\bigO{N_c^4}$ for a translational symmetry breaking
cluster to $\bigO{N_c^3}$ for a DCA cluster is therefore a crucial performance advantage that makes calculations on clusters beyond four sites
possible. The cost of the translational symmetry breaking of the DF equations introduced in Sec.~\ref{subsec:meth2}, in
comparison to the translationally invariant formalism of Sec.~\ref{subsec:meth1}, and the necessity to work with
$\tilde{G}^{(0)}_{\omega_{n}\cK\cK'\tk\sigma}$ and $\tilde{\Sigma}_{\cK\cK'\tk}$ is negligible.

The exact dual fermion equations contain not just two-particle but general $n$-particle DCA cluster vertex functions. Computing these is
achieved by a two-dimensional fast Fourier transform of complexity $\bigO{N_c^2 log(N_c^2)}$, following by the assembly of vertex function
for each unique momentum point at a complexity of  $\bigO{N_c^{2n-1}}$. The overall computational cost of a DCA cluster $n$-particle vertex
function scales therefore as $\bigO{N_c^{2n-1}}$. Each diagrammatic order with $n$-particle vertex functions gives an additional order of 
$\bigO{N_c^{n-1}}$ numerical complexity. The memory complexity is order of $\bigO{N_c^{2n-1}}$ for both DCA and DCA DF methods. This
illustrates why this method  is advantageous to the dual fermion expansion around real-space cluster DMFT, which scales with
$\bigO{N_c^{2n}}$ complexity for memory requirements and an additional $\bigO{N_c^{2n}}$ of computational complexity at each diagrammatic
order. In practice, with present-day algorithms, only two- and three-particle vertices can be computed for clusters of $N_c \leq 8$, and
only two-particle vertices for clusters with size $N_c \leq 16$.

\section{Results}
\label{sec:res}

\begin{figure}
a) \includegraphics[width=0.16\textwidth]{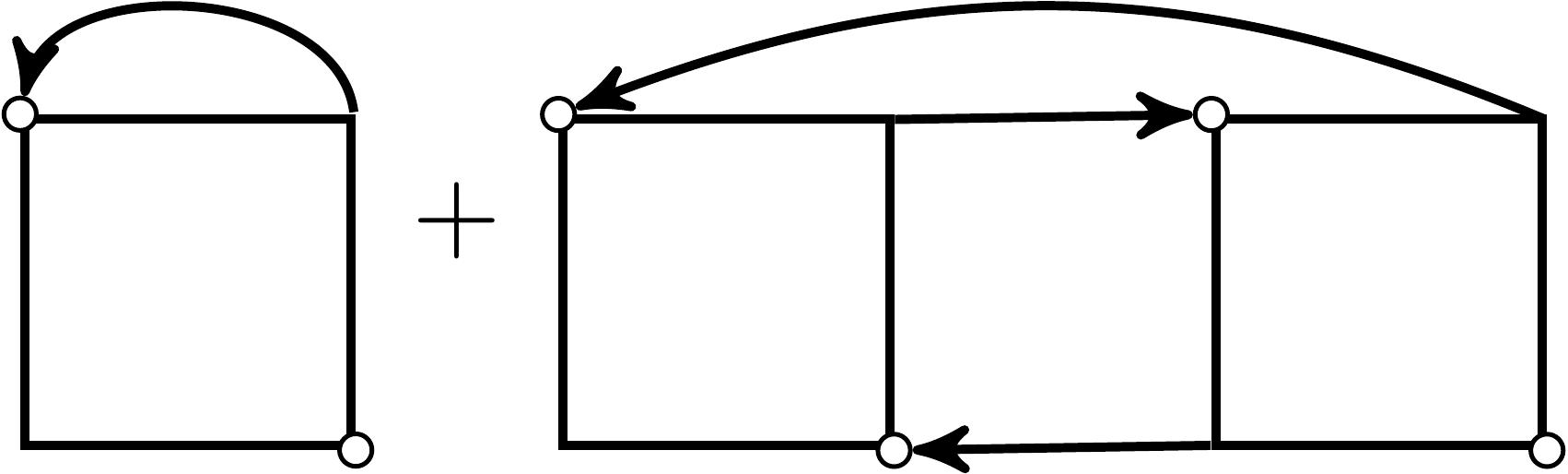}
b) \includegraphics[width=0.27\textwidth]{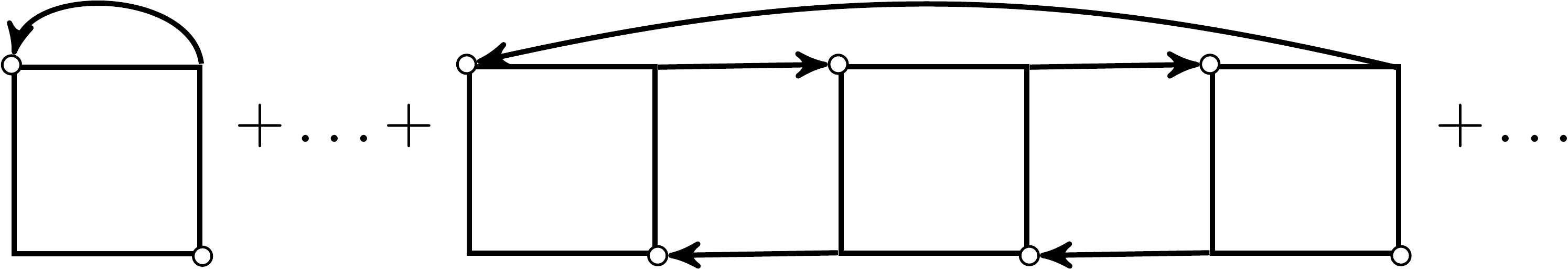}

c) \includegraphics[width=0.42\textwidth]{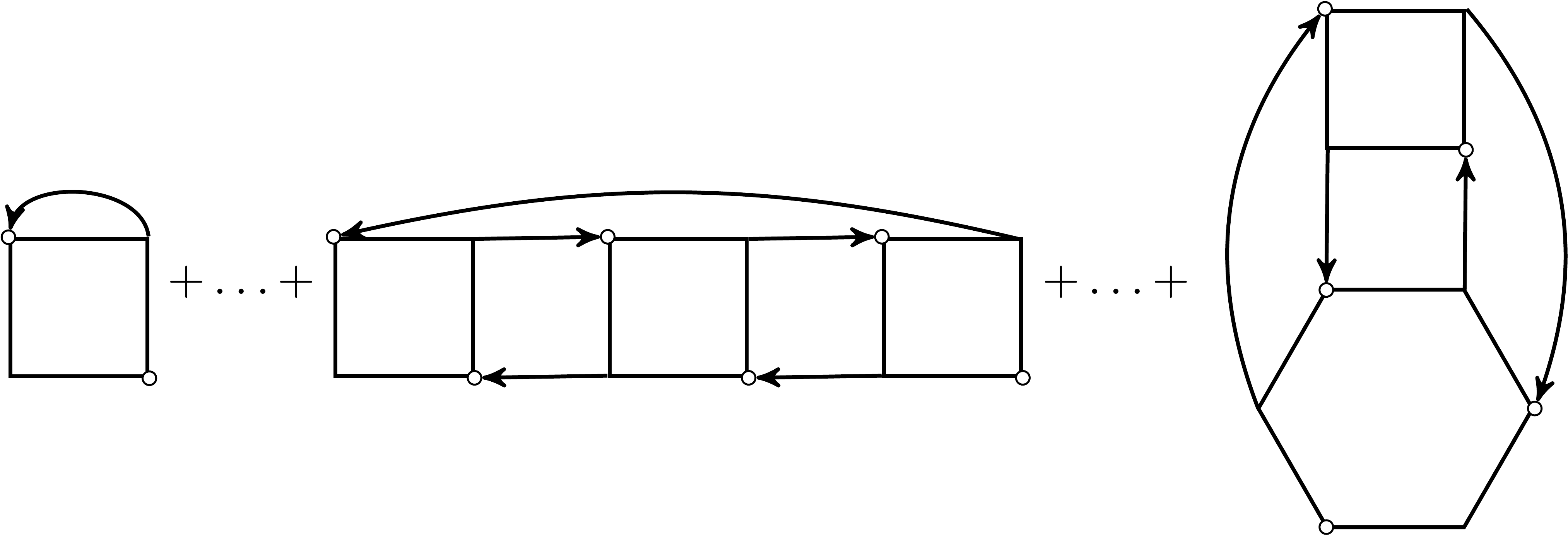}

\caption{The diagram types for the dual self-energy with major contribution. The white squares and hexagons correspond to  two- and
three-particle impurity vertex respectively. Lines correspond to the bare dual Green's function.}
\label{diagrams}
\end{figure}

In this section we present results from the methods discussed in the preceding sections. In particular, we present results obtained by the
single-site DF method,\cite{PhysRevB.77.033101} by DCA of clusters of size $N_c=4$ and $N_c=64$, and by the two cluster DF methods described
in this paper: DCA DF of Sec.~\ref{subsec:meth1} and momentum-space cluster DF method of Sec.~\ref{subsec:meth2} We choose a single set of
parameters; $U/t = 4.0$, $T/t = 0.5$ and half-filling. In this regime, the self-energy shows momentum dependence that can be resolved
accurately by large cluster DCA and several other numerical methods.\cite{1367-2630-8-8-153,VANHOUCKE201095} Data for this point is also
available in the Hubbard model benchmark.\cite{PhysRevX.5.041041}

Our DCA code uses the continuous-time auxiliary field \cite{RevModPhys.83.349,0295-5075-82-5-57003,PhysRevB.83.075122}
(CT-AUX) method with an adaptation of nonequidistant fast Fourier transforms\cite{1742-6596-402-1-012015,PhysRevLett.109.106401} to compute
vertex functions. The implementation is based on the core libraries\cite{Gaenko2017235} of the open source ALPS\cite{1742-5468-2011-05-P05001} package;
self-consistent ladder single-site DF calculations were performed by open-source OpenDF code.\cite{ANTIPOV201543}

In our cluster DF analysis, we employed three different flavors of the dual fermion approximation; all of them are in common use. First, a second-order summation of the
dual diagram. In this approach, all terms to second order in the cluster impurity two-particle vertex are included, but both higher order
terms with two-particle vertices and any contribution from higher order vertices (three-particle, four-particle, etc) are neglected. The dual
self-energy diagram of this method is illustrated in Fig.~\ref{diagrams}a. Second, we employed a ladder summation of the dual particle-hole
ladder, which contains an infinite number of diagrams of two-particle vertex functions but neglects some of the diagrams with three (or
more) two-particle vertices as well as any diagram with higher order vertices. These diagrams are illustrated in Fig.~\ref{diagrams}b.
Finally, we computed the three-particle vertex (a quantity of five momenta and frequencies) and the lowest order contribution from this
diagram. For an illustration see Fig.~\ref{diagrams}c.

\begin{figure}[tbh]
\includegraphics[width=0.49\textwidth]{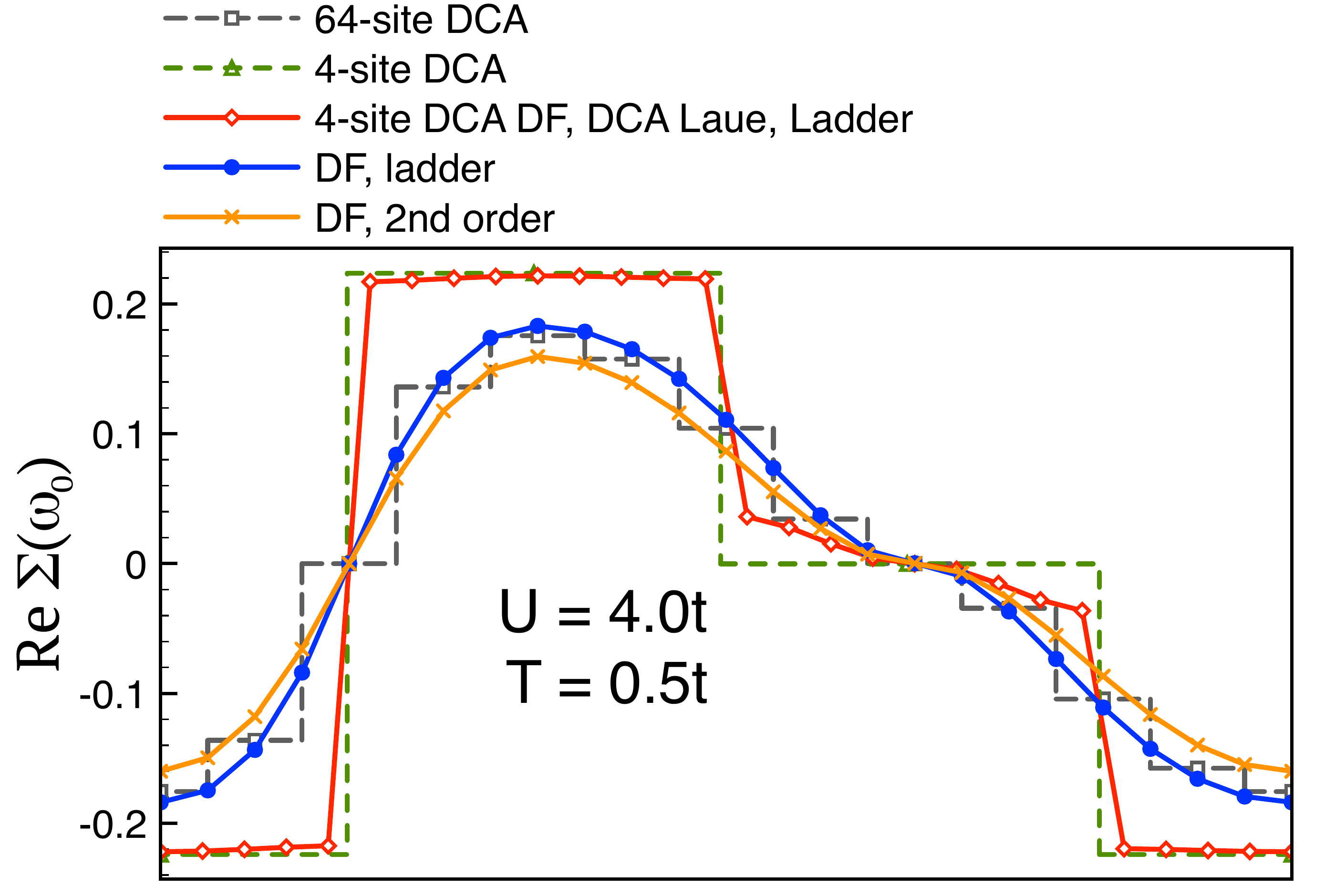}
 \includegraphics[width=0.49\textwidth]{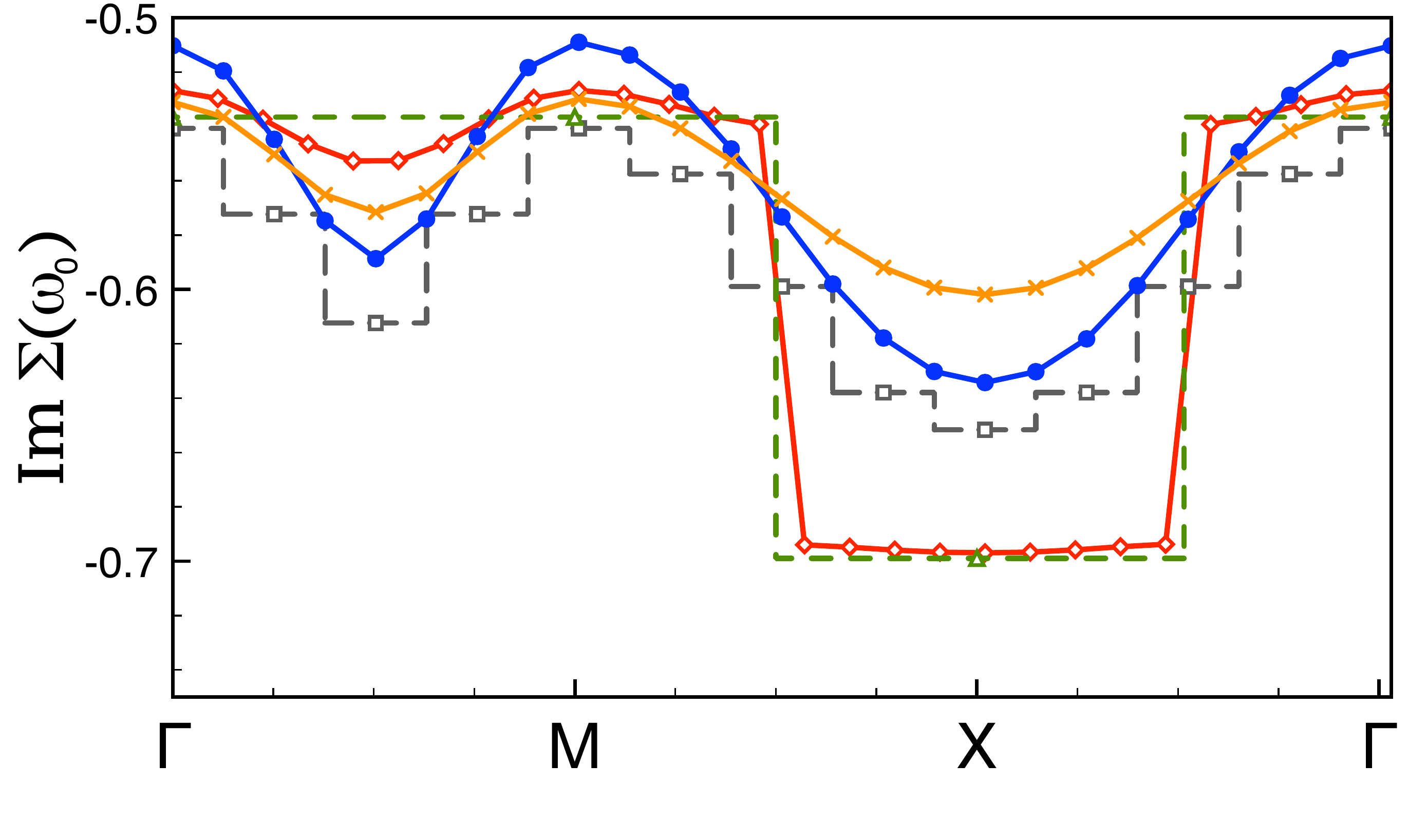}
\caption{Comparison of the real (top panel) and imaginary (bottom panel) part of the self-energy at the lowest Matsubara frequency obtained
from the different approaches. Red line with open diamonds: 4-site DCA cluster ladder DF approximation with the Laue approximation of
Sec.~\ref{subsec:meth1} (DCA DF, DCA Laue, Ladder). Blue line with closed circles: single-site Ladder DF. Orange line: single-site second
order DF. Dashed green line: $N_c=4$ DCA. Dashed gray line: $N_c=64$ DCA.}
\label{DFSE}
\end{figure}

\subsection{Dual Fermions with DCA Laue approximation}
Fig.~\ref{DFSE} shows results from 64-site DCA, 4-site DCA, single-site dual fermions in the ladder  and second-order approximation, and
4-site DCA DF with the Laue approximation as originally derived in Ref.~\onlinecite{PhysRevB.84.155106} and described in
Sec.~\ref{subsec:meth1}. Shown are the real $Re\Sigma(\omega_0)$ and imaginary $Im \Sigma(\omega_0)$ parts of the Matsubara self-energy at
the lowest Matsubara frequency $\omega_0=\pi T i$. At this point the data exhibit the largest momentum variation. The large cluster DCA
result ($N_c=64$, open square symbols) is essentially a coarse-grained version of the exact self-energy, and single-site DF and DCA for 64
sites agree reasonably well for the real part of the self-energy.

There is a small deviation of the single-site ladder DF results from the $N_c=64$ cluster results in the imaginary part of self-energy. It
is caused, as shown by \textcite{2017arXiv171000648R}, by three- and more particle correlations, in particular by the diagram shown in
Fig.~\ref{diagrams}.c. The DCA approximation for $N_c=4$ shows qualitatively similar trends, but the rough k-space discretization does not
allow to resolve details of the self-energies. Methods like cumulant, self-energy, or Green's function interpolation could be used to obtain
smoother data.\cite{PhysRevB.88.115101,PhysRevB.80.064501,PhysRevB.66.075102,PhysRevB.85.035102} Four-site ladder DF (based on the four-site
DCA results) of Sec.~\ref{subsec:meth1} show only small deviations from the 4-site DCA results, and while these results exhibit some
additional features (see in particular the real part near $X$ point), the accuracy of the solution, as compared to the large cluster
results, is generally worse than what is found in the single-site dual fermion approximation. Thus, we find that for $N_c=4$ the cluster DF
method with Laue approximation of Sec.~\ref{subsec:meth1} does not provide significant corrections to the DCA results.

\subsection{Approximations in DCA DF}
The lack of precision of the results in Fig.~\ref{DFSE}, obtained by the DCA DF method with Laue approximation of Sec~\ref{subsec:meth1}, is
due to at least one of the three approximations performed in that version of cluster dual fermions. First, the truncation of the dual
self-energy $\tilde{\Sigma}$ to an infinite series of bare ladder diagrams. Second, the neglect of diagrams with higher order vertices. And
third, the approximation of the Laue function.

\begin{figure}[tbh]
\includegraphics[width=0.49\textwidth]{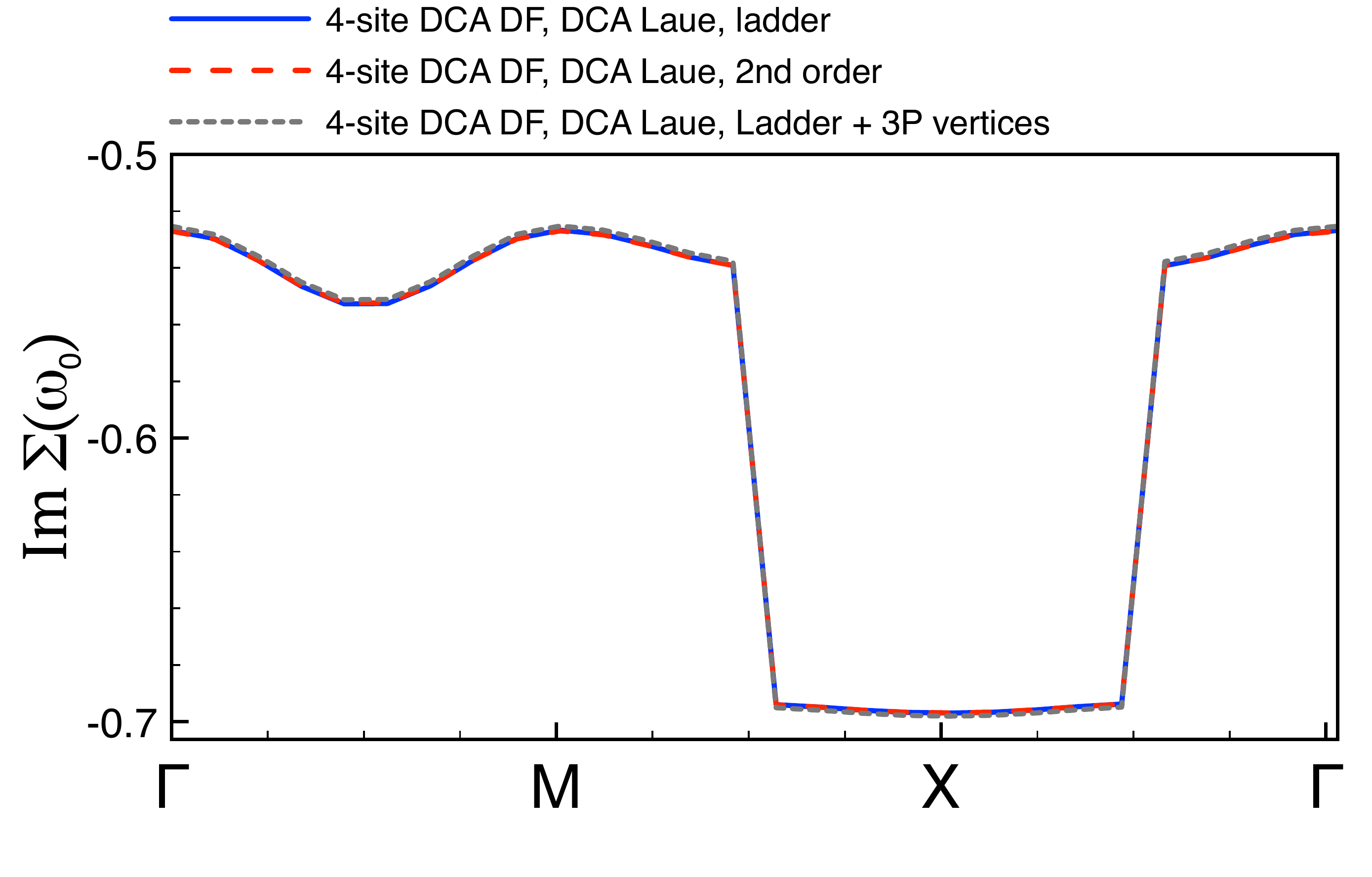}
\caption{Comparison of the different diagram topologies contribution into the imaginary part of the self-energy at the lowest Matsubara frequency. Solid blue
line corresponds to 4-site DCA cluster ladder DF results with the DCA Laue approximation, the short dashed gray line shows additional
corrections by including three particle vertex function, the long dashed red line corresponds to 2nd-order DF  with the DCA Laue approximation.}
\label{Gamma6}
\end{figure}

For the parameters considered, the difference between geometrically summed ladder and second-order diagrams is small. This is shown in
Fig.~\ref{Gamma6} as the difference between the red (long dashed) and (solid) blue lines. Two-particle diagrams that are neither of
second-order nor contained in the ladder series may in principle contribute. However, our experience with diagrammatic Monte Carlo
for dual fermions\cite{PhysRevB.94.035102} (for later related work see also Ref.~\onlinecite{PhysRevB.96.035152}) shows that in that case,
deviations between ladder and second order results are usually visible. This leads us to believe that the majority of the discrepancy comes
from a source other than diagrams with two-particle vertices.

The grey (short-dashed) line in Fig.~\ref{Gamma6} shows that contributions of the leading order three-particle vertex diagram (of
Fig.~\ref{diagrams}.c) are also small. With techniques described in Ref.~\onlinecite{2017arXiv171000648R}, it has recently become possible
to compute these corrections. Calculations employ a CT-AUX \cite{0295-5075-82-5-57003} solver with sub-matrix updates
\cite{PhysRevB.83.075122}, using a two-particle formalism developed for Raman\cite{PhysRevLett.109.106401}, NMR\cite{10.1038/ncomms14986},
and other susceptibility measurements.\cite{PhysRevLett.115.116402} While such calculations are still very expensive and only possible for
clusters up to size $8$, they are accurate enough to exclude 3-particle vertex contributions as the dominant correction to the exact result.

This implies that, at least for the small cluster studied here, the Laue approximation is the dominant source of deviations of the method of
Sec.~\ref{subsec:meth1} from the exact result. This can be explained by the fact that DCA Laue approximation neglects the inter-cluster
momentum transfer, thereby truncating the correlation length. Since anti-ferromagnetic fluctuations in this weak coupling regime are rather
long ranged, this approximation does not recover them accurately.

\subsection{Approximation-free Dual fermion DCA results} 
Section \ref{subsec:meth2} describes our formulation of DCA dual fermions, which avoids the Laue approximation altogether, at the cost of
breaking translational invariance. Thus, if all diagrams to infinite order are summed, the method converges to the exact result.

To apply this method to the four-site cluster we start from the definition of the non-interacting Hamiltonian. Since the dispersion is no
longer diagonal in the cluster reciprocal space, we first introduce the cluster hopping matrix in real-space cluster indices $\cI, \cJ$ and
momentum-space superlattice indices $\tk$:
\begin{align}
t_{\cI\cJ,\tk} = t \left(\begin{matrix}
0 & 1 + e^{-2i\tilde{k}_x} & 1 + e^{-2i\tilde{k}_y} & 0 \\
1 + e^{2i\tilde{k}_x} & 0 & 0 & 1 + e^{-2i\tilde{k}_y} \\
1 + e^{2i\tilde{k}_y} & 0 & 0 & 1 + e^{-2i\tilde{k}_x} \\
0 & 1 +  e^{2i\tilde{k}_y} & 1 +  e^{2i\tilde{k}_x} & 0 \\
\end{matrix}\right).
\label{HoppingMatrix}
\end{align}
Here, $\tilde{k}_x$, $\tilde{k}_y$ are the longitudinal and transverse part of the momentum vector $\tk$ respectively. The lattice dispersion can now be
determined by Fourier transform of the hopping matrix \eqref{HoppingMatrix} and is used to construct the bare dual Green's function matrix.
Using Eq.~\ref{DualAFAction} we construct the perturbation series to compute the dual self-energy to second order. We then obtain the
lattice self-energy using Eq.~\ref{LatticeSigma} and Eq.~\ref{SymmLatticeSigma},respectively. The size of the off-diagonal components of the
self-energy $\Sigma_{\cK\cK'\tk}$ (which would be zero if all diagrams were summed to all orders) is less than $0.5\%$ of the average
diagonal element.

\begin{figure}[tbh]
\includegraphics[width=0.49\textwidth]{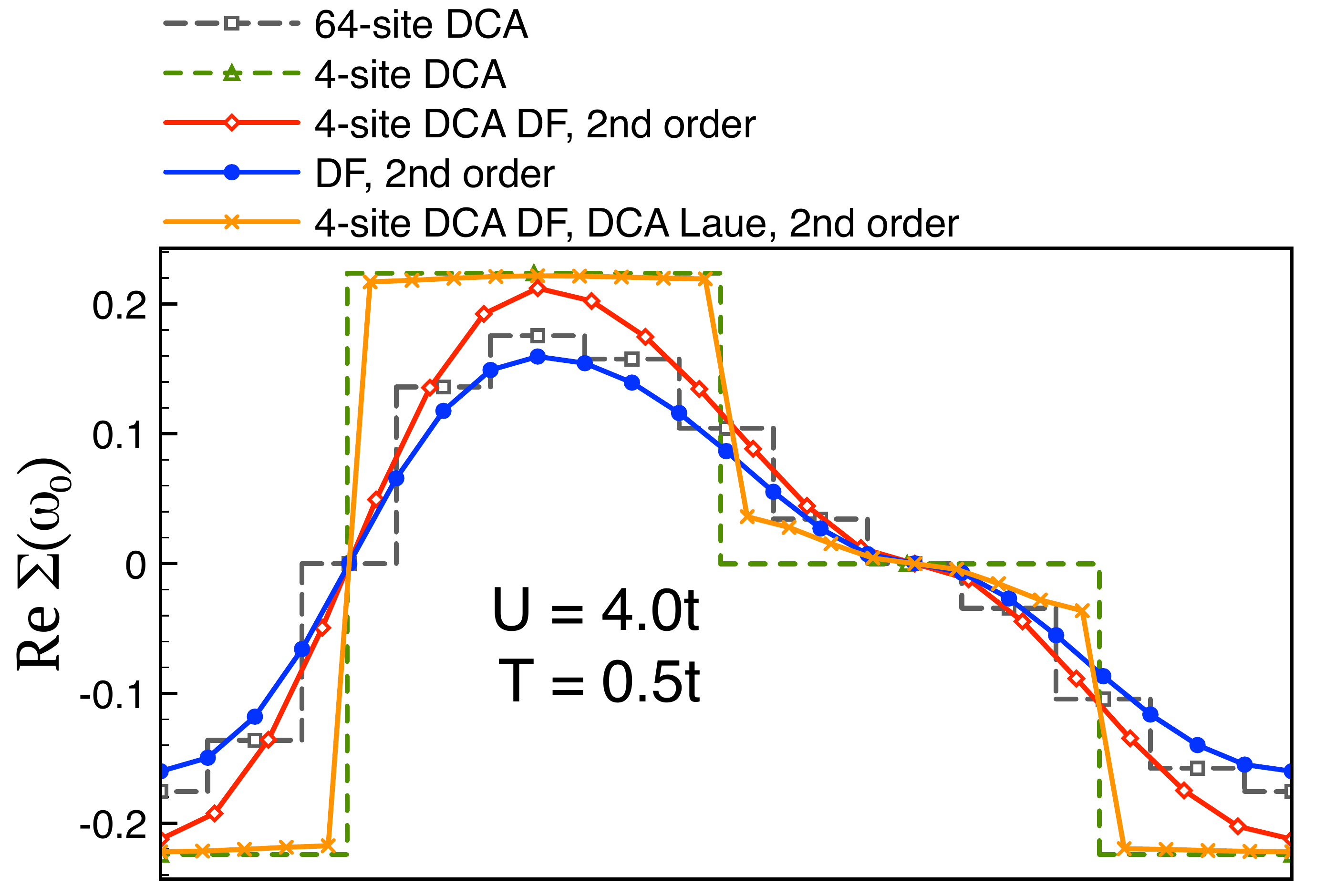}
\includegraphics[width=0.49\textwidth]{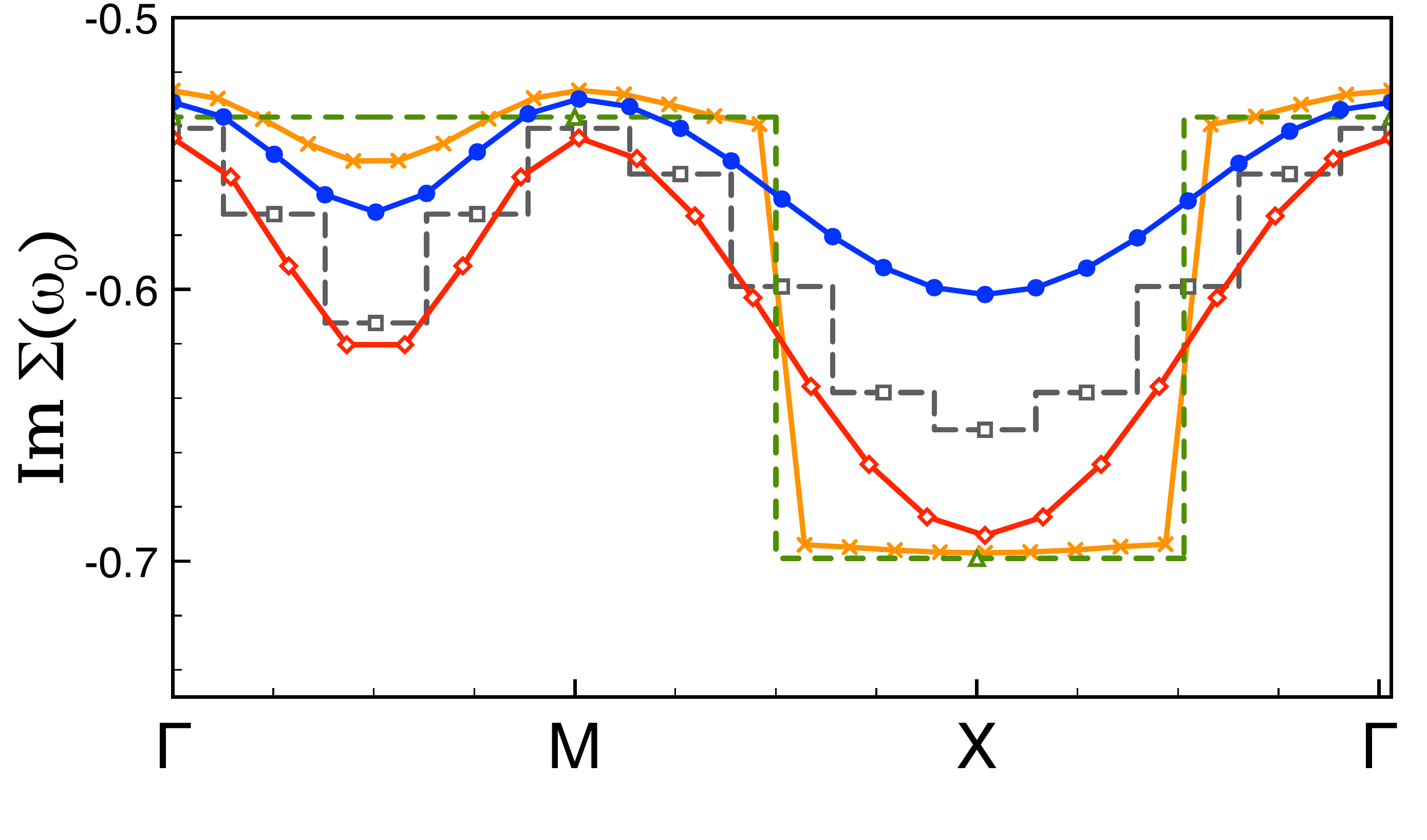}
\caption{Comparison of the real (top panel) and imaginary (bottom panel) parts of the self-energy at the lowest Matsubara frequency. Solid
red line: 4-site DCA cluster second order Dual Fermion approximation. Solid blue line: single-site second order Dual Fermions. Solid orange line:
4-site DCA cluster second order Dual Fermion approximation with the DCA Laue approximation.
Dashed green line:
4-site DCA results. Dashed gray line: 64-site DCA results.
}\label{DFSE_new}
\end{figure}

Fig.~\ref{DFSE_new} shows the results obtained with our method of Sec.~\ref{subsec:meth2} in addition to results from the methods shown in
Fig.~\ref{DFSE}. Our momentum space DCA DF results are obtained for the second order approximation to the dual fermion self-energy for of
Fig.~\ref{diagrams}.a and use a minimal cluster of size $N_c=4$ as the DCA solution. Even at the second order approximation, our cluster
dual fermion method of Sec.~\ref{subsec:meth2} improves significantly on the 4-site cluster DCA results. In particular, the imaginary part
of the local self-energy is substantially modified and much closer to the large cluster ($N_c=64$) result. This shows that basing the DF
approximation on an expansion around a cluster, which has a better representation of short range physics than a single site, makes it
possible to efficiently capture correlations that would otherwise only be contained in higher-order vertex corrections. 

\begin{figure}[tbh]
\includegraphics[width=0.49\textwidth]{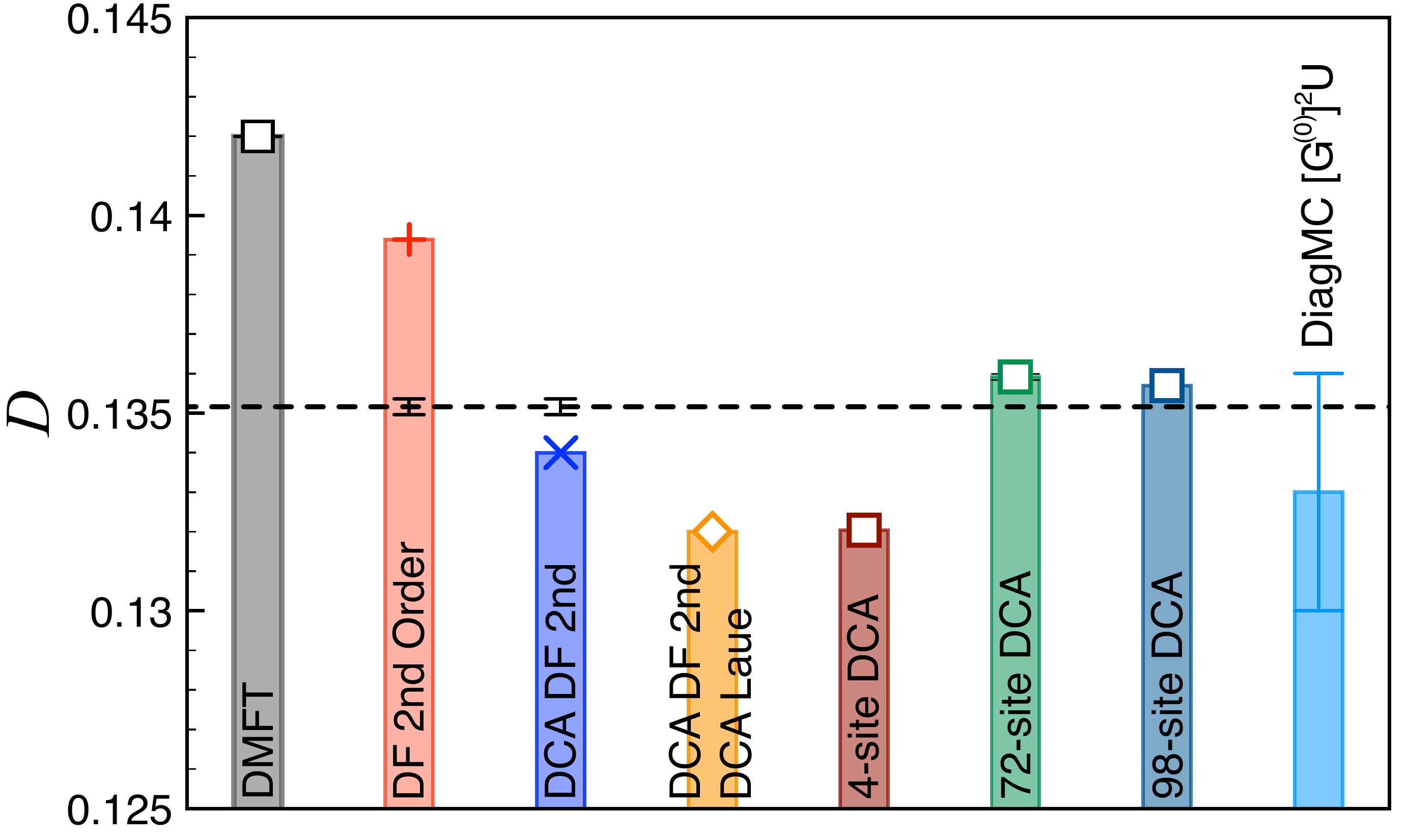}
\caption{Comparison of the double occupancy $\langle n_i n_i\rangle$ calculation obtained from DF DCA method (blue cross) in comparison to
the single-site DMFT (black square), single-site Dual Fermion method (red cross), DF DCA method with DCA Laue approximation (orange diamond)
and data from the 2D Hubbard benchmark,~\cite{PhysRevX.5.041041} the dashed black line corresponds to DCA results extrapolated to
thermodynamic limit. The symbol size for the DCA results corresponds to the stochastic error-bars.}
\label{DoubleOccupancy}
\end{figure}

In Fig.~\ref{DoubleOccupancy}, we show the results for the double occupancy from our momentum space cluster DF DCA method of
Sec.~\ref{subsec:meth2} and compare them with single-site DF (2nd order) results, results from the method of Sec.~\ref{subsec:meth1}, and
data from Ref.~\onlinecite{PhysRevX.5.041041}. The double occupancy is a two-particle quantity and directly related to the interaction
energy. DCA extrapolated  to the thermodynamic limit from a large number of clusters (dashed line) are the best reference results available 
and are, within error bars, consistent with DiagMC.\cite{PhysRevX.5.041041}

At the second order approximation shown here, DCA DF of Sec.~\ref{subsec:meth2} improves the single-site DF results by a factor of three and
the 4-site DCA data by a factor of two. The remaining discrepancy indicates that system exhibits non-local non-perturbative correlations
beyond those nearest neighbor correlations captured by the 4-site DCA cluster. Larger clusters and/or additional dual fermion diagrams will
be needed to obtain systematic convergence to the exact result.

\section{Conclusions}\label{sec:conc}
In this paper, we present a momentum-space cluster dual fermion expansion which uses the dynamical cluster approximation as a starting
point. Our method is free of approximations, as it is based on exact mapping from the original lattice model to the DF system, and converges
to the exact result if all diagrams in the dual fermion perturbation series are summed to infinite order. As a proof of principle, we show
results for the second-order DF self-energy around a four-site DCA cluster at the half-filling and demonstrate that this method is in good
agreement with reference data, and in particular provides large non-local corrections to the imaginary part of the self-energy. The avoidance
of the Laue approximation leads to a smooth convergence of the dual self-energies.

While we restrict our calculations to second-order perturbation theory in the DF diagrams, methods such as diagrammatic Monte
Carlo\cite{PhysRevB.94.035102} could be applied to recover contributions from higher order diagrams. Whether it is more advantageous in
practice to consider these contributions or to extend the simulations to larger cluster sizes is currently under investigation.

Our DF dual fermion perturbation theory builds a bridge between cluster methods (which are non-perturbative and contain all short-range
correlations) and DF methods (which are based on a perturbative summation of the non-local corrections). It will find applications wherever
the underlying DCA approximation reveals a strong momentum dependence of the self-energy. The capability of the DF formalism to yield 
continuous k-dependence for the self-energies and the vertices at affordable numerical cost will allow to make connections to experimentally
measured k-dependent phenomena such as transport and optical conductivities, where the discrete momentum structure of the DCA leads to discontinuities
in the vertex correction terms.\cite{PhysRevB.80.161105} It will also allow to study the  evolution of Fermi surfaces and the competition
between long-wavelength phenomena (such as antiferromagnetic nesting) with the strong short-range correlations that seem to be responsible
for the pseudogap\cite{PhysRevB.80.245102,PhysRevLett.114.236402} and superconductivity\cite{PhysRevLett.110.216405} in the two-dimensional
Hubbard model.

\acknowledgments{
This project was supported by the Simons Foundation via the Simons Collaboration on the Many-Electron Problem.  This work used computer
resources from the Extreme Science and Engineering Discovery Environment (XSEDE), which is supported by National Science Foundation grant
number ACI-1548562.
}
\bibliographystyle{apsrev4-1}
\bibliography{main.bib}
\end{document}